\RequirePackage{lineno}
\documentclass[aps,prc,twocolumn,superscriptaddress, showpacs]{revtex4}

\usepackage{epsfig}
\usepackage{amssymb}
\usepackage{bm}
\usepackage{graphicx,color}
\usepackage{subfig}
\usepackage{graphicx}
\usepackage{mwe}
\usepackage{dcolumn}
\usepackage{epstopdf}
\usepackage{csquotes}
\usepackage{amsmath}
\usepackage{tikz,xcolor,hyperref}
\definecolor{lime}{HTML}{A6CE39}
\DeclareRobustCommand{\orcidicon}{
	\begin{tikzpicture}
	\draw[lime, fill=lime] (0,0) 
	circle [radius=0.16] 
	node[white] {{\fontfamily{qag}\selectfont \tiny ID}};
	\draw[white, fill=white] (-0.0625,0.095) 
	circle [radius=0.007];
	\end{tikzpicture}
	\hspace{-2mm}
}
\foreach \x in {A, ..., Z}{\expandafter\xdef\csname orcid\x\endcsname{\noexpand\href{https://orcid.org/\csname orcidauthor\x\endcsname}
			{\noexpand\orcidicon}}
}

\begin{document}



\title{A Complete Approach to Determine the $^3$He neutron incoherent scattering length $b_i$}
\author{H. Lu\orcidB{}} \affiliation{Indiana University/CEEM, 2401 Milo B. Sampson Lane, Bloomington, IN 47408, USA}
\author{O. Holderer\orcidE{}} \affiliation{Forschungszentrum J\"ulich GmbH, J\"ulich Centre for Neutron Science (JCNS) at Heinz Maier-Leibnitz Zentrum (MLZ), 85747 Garching, Germany}
\author{A. Ioffe\orcidH{}} \affiliation{Forschungszentrum J\"ulich GmbH, J\"ulich Centre for Neutron Science (JCNS) at Heinz Maier-Leibnitz Zentrum (MLZ), 85747 Garching, Germany}
\author{S. Pasini\orcidF{}} \affiliation{Forschungszentrum J\"ulich GmbH, J\"ulich Centre for Neutron Science (JCNS) at Heinz Maier-Leibnitz Zentrum (MLZ), 85747 Garching, Germany}
\author{P. Pistel} \affiliation{Forschungzentrum J\"ulich GmbH, ZEA-1, 52425 J\"ulich Germany}
\author{Z. Salhi\orcidG{}} \affiliation{Forschungszentrum J\"ulich GmbH, J\"ulich Centre for Neutron Science (JCNS) at Heinz Maier-Leibnitz Zentrum (MLZ), 85747 Garching, Germany}
\author{B. M. Goodson\orcidC{}}
\affiliation{School of Chemical and Biomolecular Sciences, Southern Illinois University, Carbondale, IL 62901, USA}
\author{W. M. Snow\orcidD{}}
\affiliation{Indiana University/CEEM, 2401 Milo B. Sampson Lane, Bloomington, IN 47408, USA}
\author{E. Babcock\orcidA{}}
\email[]{e.babcock@fz-juelich.de}
\affiliation{Forschungszentrum J\"ulich GmbH, J\"ulich Centre for Neutron Science (JCNS) at Heinz Maier-Leibnitz Zentrum (MLZ), 85747 Garching, Germany}

\date{\today}

\begin{abstract}
We report the first results from a new approach for measuring the $^3$He neutron incoherent scattering length $b_{i}$. $b_{i}$ is directly proportional to the difference $\Delta b=b_{+}-b_{-}$ in the two low-energy s-wave neutron-nucleus scattering amplitudes $b_{+}$ and $b_{-}$, corresponding to the singlet $J=0$ and triplet $J=1$ states of the neutron-$^3$He interaction, respectively. An accurate measurement of $b_{i}$ can help distinguish among different models of three-nucleon interactions by comparison to {\it ab initio} nuclear theory calculations. The neutron birefringence caused by $\Delta b$ results in neutron spin rotation around the nuclear polarization. We measured $\Delta b$ using polarized neutron spin rotation and the transmission of neutrons through a $^3$He gas target polarized in situ by spin-exchange optical pumping. This brief test measurement, conducted at the FZ-J\"ulich neutron spin echo spectrometer at the Heinz Maier Leibnitz Zentrum (MLZ), yielded $\Delta b = [-5.27 \pm 0.05$ (stat.) $- 0.05$ (syst.)] fm. We argue that this method can be improved in precision to resolve the discrepancies between two prior measurements of $b_i$ which are dependent on the polarized absorption cross section $\sigma_p$. Further with absolute $^{3}$He polarization via NMR (in a properly-shaped cell) concurrent with accurate neutron transmission measurements, $\sigma_p$ can be measured to obtain independent values of $b_{+}$ and $b_{-}$. 
\end{abstract} 

\maketitle


Precision measurements of the scattering amplitudes in the $n+^{3}$He system provide important tests for {\it ab initio} theoretical calculations of the properties of few-nucleon systems. Three-body (3N) interactions among nucleons are now estimated to provide about 5\% of the total binding energy of stable nuclei~\cite{Atkinson2020}. The development of a global model of bound nuclei that can both explain the binding energies of stable nuclei and can also make reliable predictions out to the extremes of nuclear stability is a major long-term goal for nuclear physics, with important scientific applications for astrophysics and for our understanding of the process of formation of the heavy elements~\cite{Johnson2020}. Although theoretical models for the possible forms of nuclear three-body forces exist and give a rough estimate for the relative sizes and spin/isospin dependence of three- and higher-body effects compared to two-nucleon forces, more precise experimental data on systems with few nucleons is needed to determine the relative strengths of these forces.  The binding energies of $^{3}$H, $^{3}$He, and $^{4}$He are essential data for this purpose in the $A=4$ system and are measured with high precision. Theoretical calculations of the binding energy of $^{4}$He using the Green's function Monte Carlo technique~\cite{Kamada2001} including some using phenomenological three-nucleon interactions~\cite{Wiringa2000, Nogga2002, Lazauskas2004, Viviani2005} differ from experiment by 1\%. Theoretical analysis also shows that the information on the nuclear three-body force from the binding energy of $^{4}$He is not independent of that from three-body bound systems and is mainly sensitive to the spin-independent component of the nuclear three-body force~\cite{Pudliner1995, Machleidt2011, Carlson2015, Baroni2018}.  To better constrain the spin-dependent parts of the nuclear three-body force, data are required on the spin-dependent scattering of three- and four-body systems with precision at the sub-percent level.

Existing data on the difference $\Delta b=b_1-b_0$ of the two scattering lengths $b_{0}$ and $b_{1}$ for the two total angular momentum $J=0,1$ values of n-$^{3}$He are inconsistent. $\Delta b$ is proportional to the n-$^{3}$He incoherent scattering length $b_i$. The two best measurements of $b_{i}$ using neutron spin echo~\cite{Zimmer2002} and neutron interferometry~\cite{Huber2014} are inconsistent based on the quoted errors, as are the three measurements of the n+$^{3}$He coherent scattering length $b_{c}$ using neutron interferometry~\cite{Kaiser1979, Huffman2004, Keller2006}. Different theoretical calculations of $b_{1}$ and $b_{0}$ available for comparison at the time employing NN+3N interactions, such as the standard potential models AV18 + UIX, AV18 +UIX + V3~\cite{Hofmann2003, Hofmann2008}, and AV18 + LL2~\cite{Kirschner2010}, were also not in agreement. Improved measurements of both $b_{i}$ and $b_{c}$ are needed to help resolve the inconsistencies shown in previous work. Improved precision on both $b_{i}$ and $b_{c}$ can also help distinguish among different models of few-nucleon interactions.  The description of the $^{4}$He continuum just above the $n+^{3}$He threshold is challenging for existing theory to treat, and changes to existing 3N force models are proposed as a possible solution to existing discrepancies. Fortunately several new theoretical techniques have been developed to tackle nuclear four and five body systems~\cite{Nollett2007, Kievsky2008, Navratil2016, Lynn2016, Lazauskas2020, Flores:2022foz} including chiral effective theory~\cite{Machleidt2011, Piarulli2020}. Different {\it ab initio} calculational methods for nucleon scattering in $A=3$ systems deliver internally consistent results~\cite{Viviani2011, Viviani2017}, and the resonating group method (RGM) has been applied in the past to $A=4$ systems~\cite{Quaglioni2008, Navratil2010}. We judge that the prospects for improved theoretical calculations in the $n+^{3}$He system are good.

In general the total free n-nucleus scattering length is given by $a=a'+ia''$ where $a'$ and $a''$ are real. The imaginary term $a''$ arises from absorption, which is very large for the n-$^3$He system. For the forward scattering amplitudes measured in this work the bound scattering lengths are observed. The bound scattering length is related to $a$ by $b=a(A+1)/A$ where $A$ is the nucleus to neutron mass ratio. The two s-wave neutron-nucleus scattering amplitudes $b_{+}$ and $b_{-}$, corresponding to the total nucleus plus neutron angular momentum $J=I+1/2$ and $J=1-1/2$ scattering channels from a nucleus of spin $I$ and neutron of spin $s$, can be expressed as 
\begin{equation}
b=b_{c}+{\frac{2b_{i}}{\sqrt{I(I+1)}}}{s} \cdot {I}.
\end{equation}
The coherent scattering length is thus 
\begin{equation}
    b_{c}=\frac{(I+1)b_{+}+Ib_{-}}{(2I+1)}
\end{equation}
and the incoherent scattering length is
\begin{equation}
 b_{i}=\frac {I\sqrt{I+1}(b_{+}-b_{-})}{(2I+1)},
\end{equation}
which is directly proportional to $\Delta b=b_{+}-b_{-}$. The two values $J=0,1$ of the total spin for $I=1/2$ imply $b_+\equiv b_1$ and $b_-\equiv b_0$ for the triplet and singlet scattering lengths, respectively.

$\Delta b$ can be measured by observing the precession of the neutron spin as neutrons pass through a polarized nuclear target, named ``pseudomagnetic precession"~\cite{Baryshevsky1965} in the literature. Although this phenomenon was initially described~\cite{Baryshevsky1965, Abragam1982} in terms of a fictitious ``pseudomagnetic field" inside the medium, $\Delta b$ originates from neutron-nucleus scattering. The optical theorem~\cite{Sears} relates the spin dependence of the neutron optical potentials associated with the scattering amplitudes $b_{+}$ and $b_{-}$ to a two-valued neutron index of refraction ($n_{+}$,$n_{-}$)  depending on the relative orientation of the neutron spin and the nuclear polarization:
\begin{equation}
    \begin{split}
        n^{2}_{\pm}=1-{\frac{4\pi}{k^{2}}}N(b_{coh}+b_{\pm}), \\ \\
\Delta n=(n_+-n_-)\approx -{\frac{2\pi}{k^{2}}}N (b_{+}-b_{-}),
\label{eq:newindex2}  
      \end{split}
\end{equation}
\noindent where $N$ is the number of nuclei per unit volume, $k=2\pi/\lambda$ is the neutron wave number, and the approximation in the second expression is valid in our case as the neutron index of refraction is $\simeq 1$. $\Delta n$ makes the medium optically birefringent for neutrons so that the two helicity components of the neutron spin accumulate different phases, $kn_{\pm} d$, in the forward direction as neutrons propagate a distance $d$ through the target. Therefore neutron spins orthogonal to the nuclear polarization direction of the target precess around the nuclear polarization by an angle $\phi^*=k \Delta n d$. 

We can write the neutron precession angle $\phi^*$ created by the incoherent scattering length $b_i\propto{\Delta b}$ of the polarized $^3$He~\cite{Zimmer2002, Huber2014} as
\begin{equation}
    \label{eq:pseudoangle1/2}
\phi^{*} =-{\frac{1}{2}}\lambda P_3Nd{\Delta b}=-\frac{2 \lambda P_3Nd}{\sqrt{3}}b_{i},
\end{equation}
where $P_3$ is the $^3$He polarization, $N$=[He] is the $^3$He density, and $d$ is the neutron path length through the $^3$He. For nuclei such as $^3$He which possess a very large spin-dependent component to the neutron cross section, one can determine the constant of proportionality $\lambda P_3Nd$ using neutron measurements and write the measured quantity $\Delta b$ as follows:

\begin{equation}
\Delta b=\frac{2\phi^*}{\lambda P_3Nd}=\frac{\sigma_p}{\lambda_{th.}}\frac{{2}\phi^*}{{\rm cosh^{-1}}R} . 
\label{Req}
\end{equation}
\noindent 
Here $R$ is the ratio of unpolarized neutron transmission of polarized $^3$He, $T(P_3)$, to the transmission of unpolarized $^3$He, $T(0)$, $\sigma_{p}$ is the polarized $^3$He spin dependent neutron absorption cross section, and $\lambda_{th.}=1.798~$\AA~is the thermal neutron wavelength chosen by convention as a reference point for neutron absorption cross sections. The total n-$^3$He absorption cross section $\sigma_a=(4\pi/k)b''$ obtained from the imaginary part of $b$ by the optical theorem \cite{Sears} can be written in terms of the polarization-independent and polarization-dependent terms as:
\begin{equation}
\sigma_a=\sigma_{un}\mp P_3\sigma_p,
\label{absorption}
\end{equation}
where the sign convention $\mp$ is for $P_3$ parallel (-) or anti-parallel (+) to the neutron spin. Here $\sigma_{un}=(5333 \pm 7)$ barn is the total unpolarized neutron absorption cross section, and $\sigma_p$ can be expressed as $\sigma_{p}=(1-\sigma_1/\sigma_{un})\sigma_{un}$. Both $\sigma_{a}$ and $\sigma_{un}$ are measured to be proportional to the neutron wavelength $\lambda$ to high precision~\cite{Als-Niesen1964, Borkazov1982, Keith2004}. For an unpolarized neutron beam of $n$ neutrons with half spin up ($n^+$) and half spin down ($n^-$) neutrons the corresponding transmission of Eq.~\ref{absorption} is
\begin{equation}
    T^{\pm}={\frac{n^{\pm}}{n}}={\frac{1}{2}}{\rm exp}\left (-(\sigma_{un}\mp P_3\sigma_p){\frac{\lambda Nd}{\lambda_{th.}}}\right ).
\end{equation}
Thus, the transmission of unpolarized neutrons through polarized $^3$He is:
\begin{equation} 
T(P_3)= {\rm exp}\left (-{\frac{\sigma_{un}}{\lambda_{th.}}}\lambda Nd\right ){\rm cosh}\left ({\frac{\sigma_{p}}{\lambda_{th}}}\lambda P_3Nd\right ).
\label{TP}
\end{equation}
Since the unpolarized transmission is simply: 
\begin{equation}
    T(0)={\rm exp}\left (-{\frac{\sigma_{un}}{\lambda_{th.}}}\lambda Nd\right )
\end{equation} 
with two neutron transmission measurements giving $R=T(P_3)/T(0)$ one directly experimentally obtains the product $\frac{\sigma_p}{\lambda_{th.}}\lambda P_3Nd$ from ${\rm cosh}^{-1}R$, leading to eq. \ref{Req}. 

 Thus one still needs to determine $\sigma_{p}$. If the triplet absorption rate $\sigma_1$ were zero, so that there would be only absorption in the singlet state $\sigma_{0}$, then $\sigma_{p}=\sigma_{un}$, where $\sigma_{un}$ is known to $\simeq 0.1\%$. However the upper bound on $\sigma_p$ 
 from previous experiments is several-percent ~\cite{Passell1966, Borzakov1982} and limits the precision of this technique. There is no reason to expect that $\sigma_{1}$ is zero, as theoretical calculations show~\cite{Hofmann2003, Hofmann2008}. The work in Ref.~\cite{Zimmer2002} used an average of the experimental determinations \cite{Passell1966, Borzakov1982} to arrive at a value that can be reinterpreted as $\sigma_1=57$ barn. Conversely the work in Ref.~\cite{Huber2014} used a combination of theoretical predictions and the measured thermal absorption cross section to estimate $\sigma_1=24$ barn \cite{HuberPC}. For lack of better knowledge of $\sigma_1$, we will use the latter value in our analysis below but also present the result independent of the value of $\sigma_1$ as was also done in \cite{Huber2014} for comparison purposes.



$^{3}$He SEOP cells can also enable a measurement of $\sigma_p$. An independent 0.1$\%$ measurement of $P_3$ combined with accurate measurements of $R(\lambda)$ through an {\it in-situ} polarized $^3$He sample using the time-of-flight (TOF) method ({\it i.e.} a wavelength scanned neutron beam) as in Ref. \cite{Chupp2007} could provide a $\simeq 0.1\% $ accuracy for $\sigma_p$, allowing determination of $\sigma_1$ to $\approx 5$ barn accuracy. Several atomic physics methods have determined $P_3$ to high precision \cite{Romalis98,Babcock2005,Nikolaou14,Wilms97} so this approach should be feasible. 

In order to determine $P_3$ on a neutron beamline, we propose to use the ``self-magnetometry" of a polarized $^3$He sample in a defined shape such as in a long tube parallel or perpendicular to the applied $B_0$ field. The $^3$He magnetization $M_3=\mu_3P_3N$ will generate a magnetic field of 
\begin{equation}
B_3=\mu_0 M_3\left(1-{\frac{2}{3}}\right)=\mu_0 {\frac{M_3}{3}}
\label{parallel}
\end{equation}
when the tube's axis is parallel to $B_0$ and 
\begin{equation}
B_3=\mu_0 M_3\left({\frac{1}{2}}-{\frac{2}{3}}\right)=\mu_0{\frac{-M_3}{6}}
\label{perpendicular}
\end{equation} 
when the tube's axis is perpendicular to $B_0$ \cite{Vlassenbroek96,Romalis98}. Here the first term is the magnetization minus the demagnetization factor and the $-2/3$ term, which is the same as the field from a spherical volume, arises from the scalar contact term, meaning the $^3$He spins are non-overlapping and cannot ``see" one another so the self field must be subtracted \cite{Romalis14,Limes19}. We note the magnetic moment of $^3$He $\mu_3/h$=-16217050 Hz/T is known to the ppb level \cite{Schneider22}, and the the geometric correction factor in the field parallel case for a cell with a length to diameter ratio $\simeq 5$ is $\simeq 2\%$ and very well known \cite{Chen2006,Joseph66}.
Given at one bar pressure at 25 $^\circ$C there are $2.43\times 10^{25}$ atoms m$^3$ and the gyromagnetic ratio of $^3$He, $\gamma_3=3.24\times 10^7$ Hz/T, the product $f_3=\mu_0\mu_3P_3N\gamma_3=10.6$ Hz at $P_3=1$ and $N=1$ bar. Thus for the field-parallel case, upon a reversal of $P_3$, an NMR frequency shift of 
\begin{equation}
    \Delta f_3=2B_3\gamma_3=2\mu_0 \frac{M_3}{3}\gamma_3 =\frac{2}{3}\mu_0 \mu_3 P_3N\gamma_3\simeq 5~{\rm Hz}
\end{equation} will be observed for $P_3=0.70$ at 1 bar pressure.  Since $\lambda Nd$ of eq. \ref{TP} can be calibrated by unpolarized $T(0,\lambda)$ measurements using the well-known $\sigma_{un}$ and given that one can expect $<5$ Hz NMR linewidths, 0.1\% accuracy in $P_3$ and thus $\sigma_p$ should be attainable for normal pressures by signal averaging. This method would not require new on-beamline detection methods to be developed other than a specialized cell for the purpose. {\it In-situ} polarization of the $^3$He employing adiabatic fast passage (AFP) $^3$He flipping and a TOF neutron beamline would be preferred.

Since a recent measurement of $b_{c}$ in n$+^{4}$He using perfect crystal neutron interferometry~\cite{Haun2020} reached $10^{-3}$ precision using a technique that can be directly applied to $^{3}$He, our ideas to improve $b_{i}$ are the key additional input needed to confront theory. Therefore we intend to perform a higher precision measurement of $\Delta b/\sigma_p$ and a measurement of $\sigma_p$ to obtain an absolute value for $b_i$, which could then also approach a $10^{-3}$ precision.


We have tested an accurate method to determine the real part of $\Delta b/\sigma_p$ on the J-NSE Phoenix instrument \cite{Pasini2019} as part of a different experiment to measure the same quantity for n-$^{129}$Xe and n-$^{131}$Xe \cite{Lu2022}. Our approach builds on the pioneering work of Zimmer {\it et al.}~\cite{Zimmer2002} by taking advantage of technical improvements in neutron spin echo spectroscopy to measure neutron birefringence and by exploiting the improved time stability and improved performance of polarized $^{3}$He gas targets created using spin-exchange optical pumping (SEOP)~\cite{Walker1997}. 



\begin{figure}
\begin{center}
\includegraphics[width=20pc]{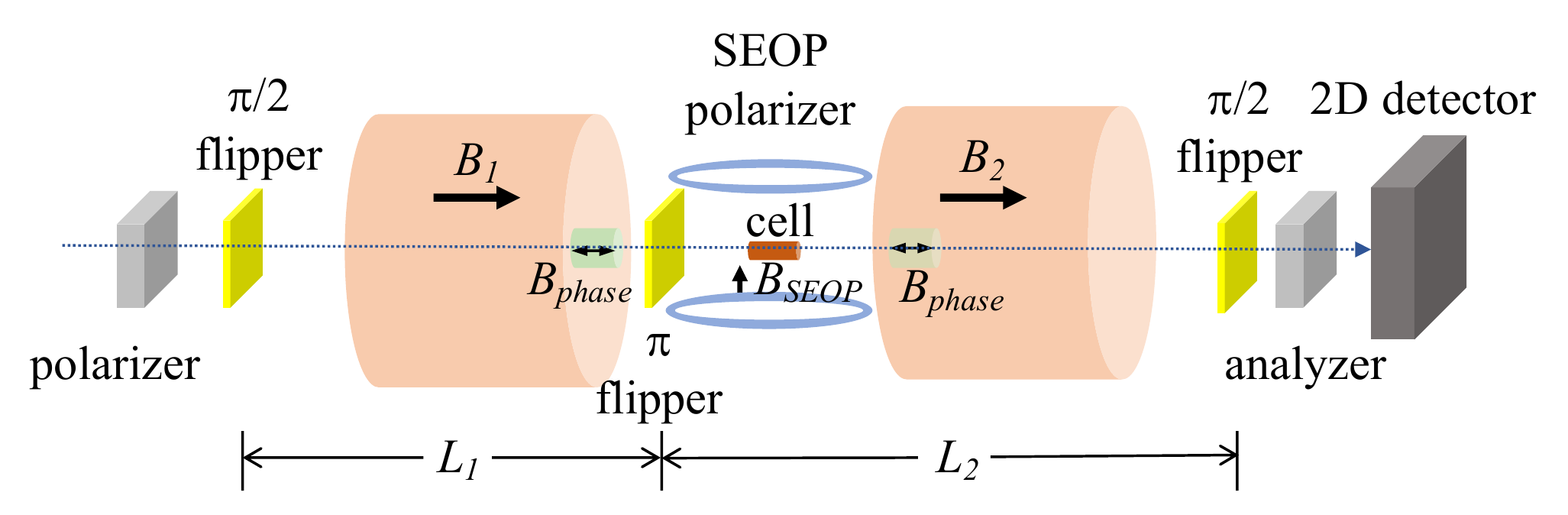}
\end{center}
\caption{\label{NSEdiagram}A schematic drawing of the J-NSE neutron spin echo spectrometer showing the coil arrangement and the SEOP-polarized $^{3}$He cell. Not shown are optical pumping lasers whose light is perpendicular to the neutron flight path and directed along $B_{SEOP}$ via a 45$^\circ$ mirror above the cell, and the oven used to regulate the cell temperature for SEOP.}
\end{figure}

Measurements of neutron birefringence in polarized nuclei were originally performed using the Ramsey method of separated oscillatory fields~\cite{Ramsey1956, Ramsey1990}. This so-called pseudomagnetic precession method~\cite{Baryshevsky1965} uses two oscillating fields before and after a solid-state nuclear-polarized sample to measure phase precessions caused by the sample. We use a variation that also uses orthogonally-precessing polarized neutrons moving through a nuclear polarized $^3$He gas sample but in a neutron spin-echo (NSE) spectrometer~\cite{Mezei1972} in order to quantify the resulting phase shifts in the neutron precession. NSE is similar to spin echo in NMR~\cite{Hahn1950} but the neutron spin is precession-encoded in space along the path of the traveling beam as opposed to in time with static spins as in NMR spin echo.

In a NSE spectrometer, polarized neutrons are first flipped by $\pi/2$ to induce precession in the orthogonal plane, they then pass through a high-field flight path with an over 1 T$\cdot$m field integral encoding a large number of spin precessions; this step is followed by a $\pi$ flip reversal of the neutron polarization in the middle of the instrument and then by a second high-field flight path identical to the first to decode the spins. The sample is typically near the middle either before or after the $\pi$ flipper and the additional precession it creates can be quantified by matching it to the precession in additional phase (compensation) coils placed around the neutron flight path. Thus the NSE method is like the Ramsey technique, but the addition of the central $\pi$ flipper allows the sample-induced phase shift to be quantified by DC phase coils rather than phase matching an oscillating RF field directly. NSE has the benefit that, because the phase coils have field integrals accurate to nT$\cdot$m compared to total instrument field integrals of 1 T$\cdot$m or more, it can encode the spins very precisely and measure very small changes in the neutron precession~\cite{Pasini2019}. 

A schematic diagram of a NSE spectrometer is shown in Fig.~\ref{NSEdiagram}. The nuclear spins of the $^3$He sample were polarized in-situ using spin-exchange optical pumping~\cite{Walker1997} in the usual sample area of the NSE spectrometer after the $\pi$ flipper. The $B_{SEOP}$ field oriented perpendicular (vertical) to the neutron flight path and the fields $B_1$ or $B_2$ of the NSE spectrometer itself, which are longitudinal (i.e. horizontal; see Fig. 1). 

Because of the work on Xe isotopes the gas target was polarized in-situ to maintain a steady nuclear spin polarization. For Xe this was mandatory because of their comparatively short $T_1$ polarization lifetimes compared to the duration of a typical NSE scan ($\simeq$20 min), especially $^{131}$Xe where $T_1<30$ s. For the $^3$He experiment this also turned out to be advantageous by decoupling time-dependent instrumental drifts from changes in $P_3$. Using a home-built NMR system for free-induction decay detection, we were able to determine that fractional changes to $P_3$ were below $0.3\%$ during our NSE measurements. The in-situ polarization equipment~\cite{Salhi2014} also enabled on-beam AFP flipping of $P_3$ during continuous pumping. The continuous polarization allows any time-dependent phase drifts in the NSE spectrometer to be fit as a time-dependent background, and the AFP flipping eliminates systematics due to possible non-perfect neutron spin-flips or non-adiabatic transport of the neutron polarization through our setup.

A 5 cm diameter cylindrical $^3$He SEOP cell made of GE180 glass with about 0.4 bar of $^3$\/He was used~\cite{Salhi2014}. This cell has somewhat rounded ends with a path length of 4.8 cm through its center. Ref.~\cite{Salhi2019} describes the SEOP instrumentation used to polarize the $^3$\/He spins. In contrast to that neutron polarizer device, here the vertical magnetic field for SEOP was provided by a set of 70 cm diameter Helmoltz coils for added flexibility and to satisfy space constraints on the J-NSE Phoenix instrument; all other components such as the lasers, NMR devices, and controls were taken directly from the device in~\cite{Salhi2019}. High-fidelity data was obtained for a 6 cm$^2$ area in a half-circle in the neutron-illuminated central portion of the cell where the path length is approximately uniform.

\begin{figure}
\centering
\includegraphics[width=0.45\textwidth]{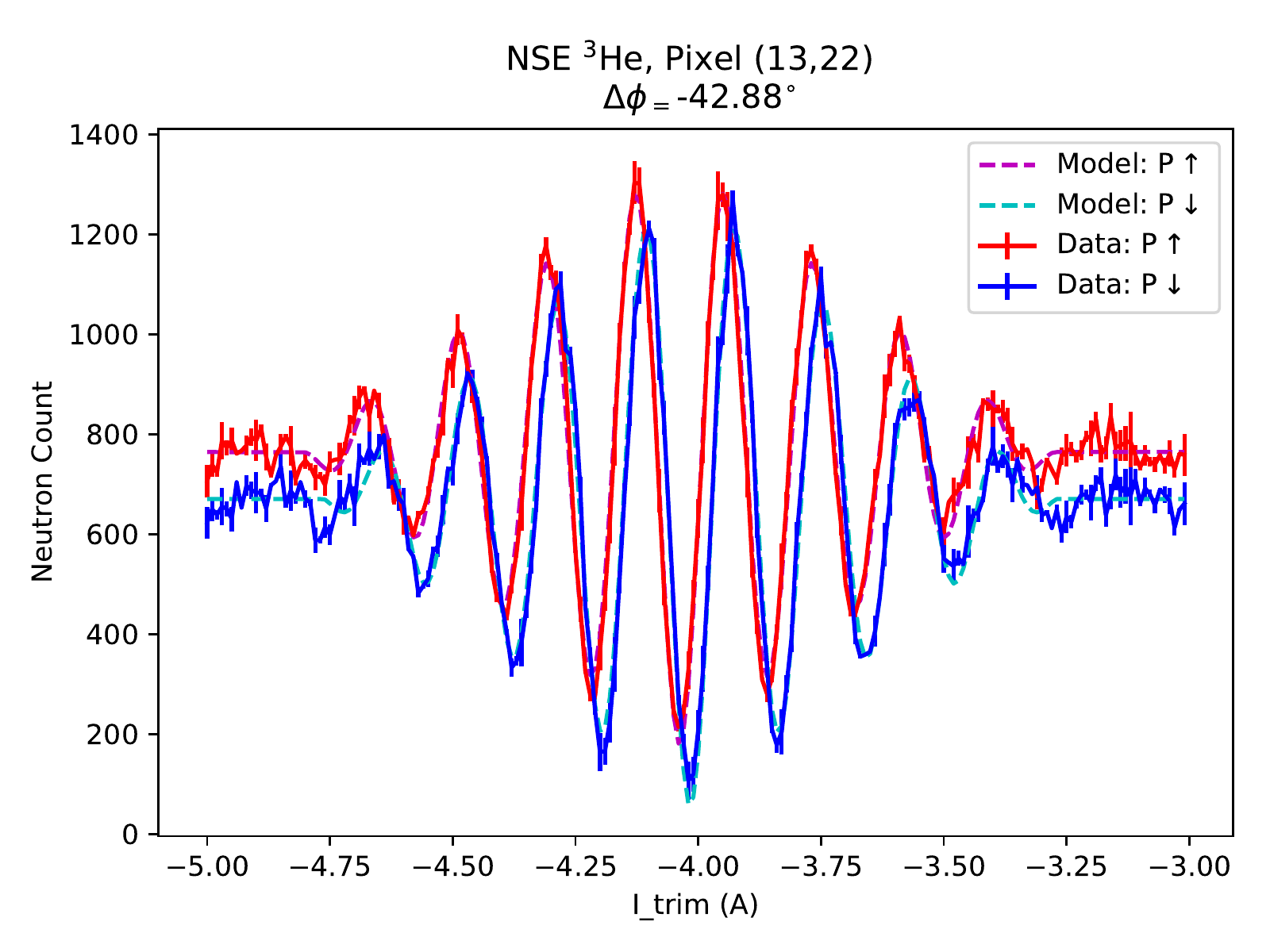}
\caption{Spin echo signals from the $^{3}$He target versus the current in the phase coil ($I_{trim}$ )for one pixel of the neutron detector. The two NSE profiles correspond to $P_3$ parallel and antiparallel to $B_{SEOP}$.} 
\label{NSEscan}
\end{figure}

A typical NSE scan is made by measuring the amplitude of the neutron polarization vector as the phase-coil is scanned in small steps around the point where the two NSE precession regions are balanced in field integral.  This action produces a spin echo envelope that shifts in proportion to precession angle $\phi^*$ of the sample. The J-NSE Phoenix spectrometer employs a position-sensitive detector allowing independent determination of $\Delta b$ for approximately each 0.5 cm $\times$ 0.5 cm region of the $^3$He cell, which can then be averaged. This practice eliminates problems that might otherwise arise from a non-uniform neutron path length though the $^3$He cell. A pair of NSE scans for the two states of the $^3$He polarization from one such pixel is shown in Figure~\ref{NSEscan}.

The NSE signal $I(B_{1})$ detected transmitted intensity after the neutron polarization analyzer as a function of the phase coil field $B_{1}$. In the expression:
\begin{equation}
    I(B_{1})=I_0[1-p\int d\lambda f(\lambda) cos(\phi_{1}-\phi_{2})],
\end{equation}
$\phi_1$ and $\phi_2$ are the precession angles in the first and second precession coils, respectively, $f(\lambda)$ is the neutron wavelength distribution, and $p$ measures the loss of contrast of the interference pattern from neutron polarization efficiencies. The instrument records the neutron intensity in each detector pixel as a function of the current of the phase coil. The neutron wavelength distribution transmitted by the velocity selector on this beamline is well fit by a triangular function, so the resulting NSE signal is this form convoluted with a cosine function which is used to fit the data for analysis.

\begin{figure}
\centering
\includegraphics[width=0.45\textwidth]{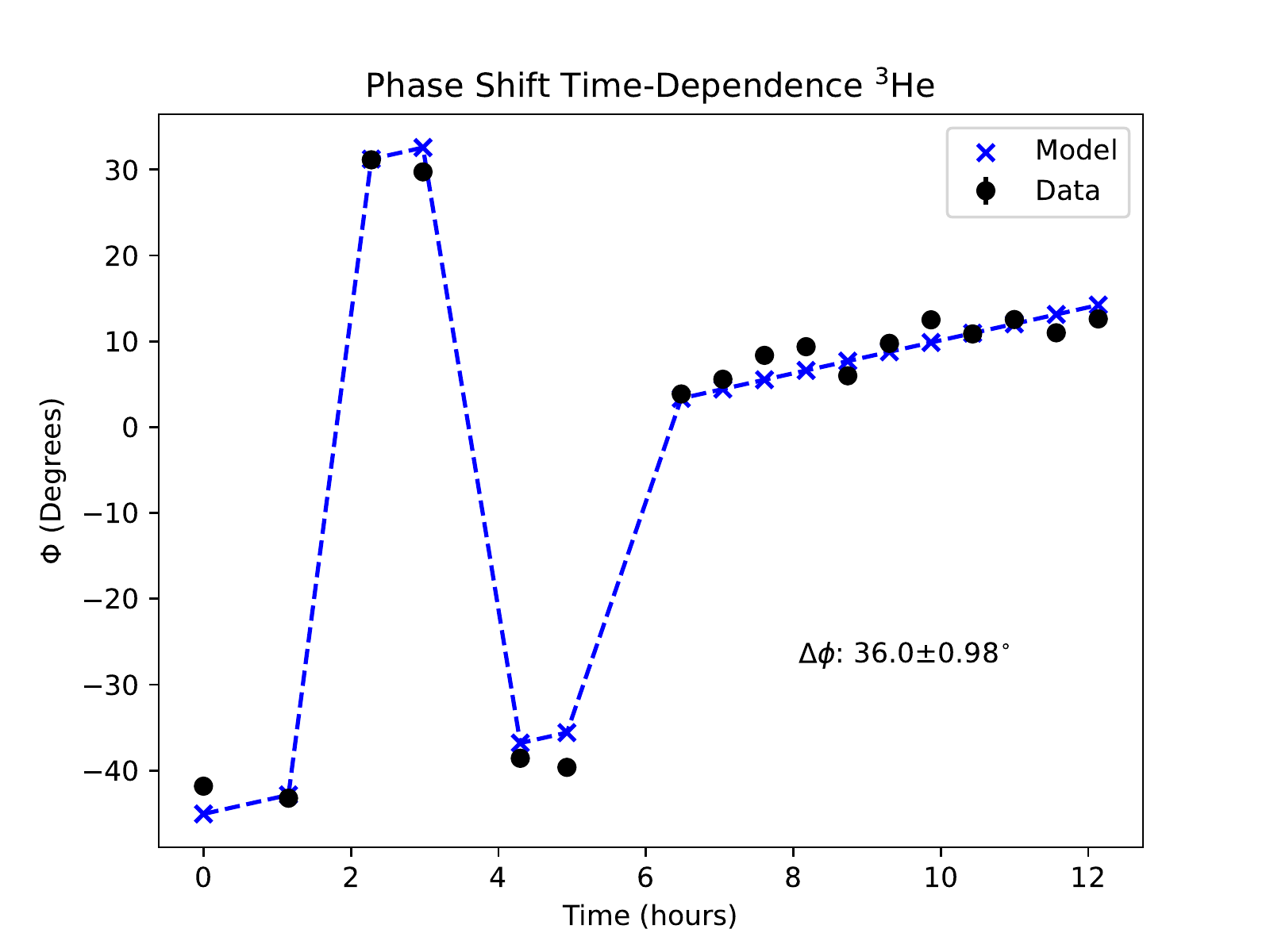}
\caption{Phase data from one pixel of the NSE detector over the entire experiment. $b_i$ values from 36 pixels representing the center of the $^3$He cell were averaged for the final result.} 
\label{3Hedata}
\end{figure}

The $^3$He $b_i$ data reported here was obtained to verify the method for the Xe measurement \cite{Lu2022}, and thus time limited. $^3$He was measured 6 hours each for polarized and unpolarized $^3$He. The data was taken in the following pattern: 2 NSE scans with $P_3$ positive; two scans with $P_3$ flipped to the negative state; two in the positive state; and 6 hours of scans with $P_3$=0. Although not needed for the determination of $b_i$, $P_3=70.6 \pm 1.6\%$ was determined using $R$ and a value of [He]=$0.3556 \pm 0.001$ bar from a separate transmission measurement using neutron TOF on the FIGARO instrument~\cite{Campbell2011}. A graph of the phase versus time for one of the pixels is shown in Fig.~\ref{3Hedata}. This data was then fit for to a step function with a linear time-dependent background to determine the measured shift $\Delta\phi=2\phi^*$ for $+P_3$ to $-P_3$ from each pixel. 

The $R$ measurement was performed by taking the weighted average of the mean intensity value of the NSE signal during the NSE scans for the $+P_3$ and $-P_3$ states to determine $T(P)$, and the mean intensity of the unpolarized $^3$He NSE scans to determine $T(0)$. This process was also performed for each pixel of the NSE scan over our region of interest to account for any variations in $d$ resulting from the cell's shape and alignment in the neutron beam. Using the mean value compensates for the  small difference in transmission of the positive versus negative $P_3$ due to a small residual neutron polarization along $B_{SEOP}$.  

The measured $\Delta\phi$ value could have a small correction due to the magnetic dipole field of the polarized gas, causing an extra precession signal in the neutron phase that is also proportional to $P_3$. This effect was discussed in Ref.~\cite{Zimmer2002}, however that work did not account for the scalar-contact term that one must include for real particles~\cite{Romalis14,Limes19}, which is different than one would expect from the simple classical result where it is assumed that one can have non-interacting and overlapping point-like particles. For n+$^3$He the contact interaction should be 0 for any reasonable first order approximations. Therefore along their flight path neutrons do not sample the classical field of the individual polarized $^3$He nuclei inside the cell. The field experienced will consist of the long range dipole fields caused by a non-spherical geometry inside the polarized cell and, since the neutron precession also integrates the field along their entire flight path, the classical field of a polarized (magnetized) volume outside of the cell. For the geometry of our experiment this would lead to a correction $<0.07\%$ and is not relevant here.  However the general arguments of the fields experienced by nonzero spin particle beams through polarized volumes is of interest for precision measurements so we include a discussion of this effect in a supplement \cite{supplement}. 

For neutron velocity selector instruments there is a small correction for the triangular wavelength distribution caused by the wavelength-dependent attenuation through the $^3$He target. Using the arguments described in the O. Zimmer work \cite{Zimmer2002} this leads to a negligible correction factor of $1.0003$ for our case because of the relatively small $Nd$ of the cell used. This correction would not be needed for an instrument that uses TOF to determine the transmitted spectrum. Our global detector count rates were $1\%$ or less of the total detector deadtime of 400 ns (e.g maximum count rate of 2.5 MHz) so detector deadtime corrections are also negligible.

Using this analysis we arrive at the final value of $\Delta b = [-5.27 \pm 0.05$ (stat.) $- 0.05$ (syst.)] fm using the values of $\sigma_{un}=5333 (7)$ barn and an estimated $\sigma_p=\sigma_{un}-\sigma_{1}=5309$ barn. From eq. \ref{eq:pseudoangle1/2} $\Delta b=4b_i/\sqrt{3}$ gives the $^3$He neutron incoherent scattering length. Rewriting the result to be independent of $\sigma_p$ as in  \cite{Huber2014,HuberPC} we obtain: 
\begin{equation}
\frac{\Delta b}{\sigma_p} =(-9.93\pm 0.09({\rm stat.})-0.09({\rm syst.}))\times 10^{-4}~{\rm \frac{fm}{b}}. 
\end{equation}
This value compares to $\Delta b/\sigma_p=(-10.1929\pm-.0760)\times 10^{-4}$ fm/b for the work of Ref.~\cite{Huber2014} and $\Delta b/\sigma_p= (-10.3628\pm-.0180)\times 10^{-4}$ fm/b for the work of Ref.~\cite{Zimmer2002}. This preliminary data shows we can readily obtain our target of $10^{-3}$ precision for $\Delta b/\sigma_p$ with a dedicated measurement. The value using this basic method is free of systemics to our knowledge. The variation of the preliminary result above with respect to previous measurements could be attributed to slow experimental drifts and that we only reversed the $P_3$ one time. The cell used had non-uniform neutron flight path and a minor shift in the cell position could lead to a one-sided error, estimations from shifting the center position of the data result in the given systematic error here. 

We implemented improvements to the measurement technique pioneered by~\cite{Zimmer2002} such as in-situ polarization of the $^3$He gas and the ability to reverse the $^3$He polarization using AFP. The in-situ polarization approach decouples the measured $\phi^*$ from time-dependent drifts that could falsely correlate with $P_3$, and prevents possible inconsistencies induced by removal and replacement of the $^3$He cell. The position-sensitive determination of the cross section reduces possible path length errors, which could be further reduced by using a flat-windowed $^{3}$He SEOP cell to eliminate variations in path-length across the beam over time, and the AFP flipping cancels errors from small residual longitudinal neutron polarization. Use of a TOF NSE instrument such as the SNS-NSE \cite{SNSNSE} will eliminate the neutron velocity selector correction. By increasing the measurement time to 1 week and using a cell with an optimized $Nd$ to minimize error propogation from the cosh$^{-1}(R)$ term we estimate one could reach a statistical accuracy of $<0.1\%$ (or 0.005 fm). Previous work~\cite{Musgrave2018} shows that the transmission measurements needed to measure the proportionality factor between $\phi^{*}$ and $\Delta b/\sigma_p$ can indeed be conducted with the required precision. With the additional measurement of $\sigma_p$ to a comparable precision, a total $10^{-3}$ precision on $b_i$ for $^3$He can be attained. 

\begin{acknowledgments}
H. Lu and W. M. Snow acknowledge support from US National Science Foundation (NSF) grants PHY-1913789 and PHY-2209481 and the Indiana University Center for Spacetime Symmetries. H. Lu received a Short-Term Grant, 2019 no. 57442045 from DAAD the German Academic Exchange Service. B.M. Goodson acknowledges support from the NSF (CHE-1905341), DoD (W81XWH-15-1-0272, W81XWH2010578), and a Cottrell Scholar SEED Award from Research Corporation for Science Advancement. P. Guthfreund (ILL) and K Zhernenkov performed a calibration measurement of $N$ ({\it i.e} [He]) on FIGARO \cite{Campbell2011} aiding this work. We acknowledge G.M. Schrank for discussions and M. Huber for detailed discussions of NIST work on $b_i^3$ and estimates of $\sigma_1$ for $^3$He.
\end{acknowledgments}


\section{Supplement}

As described in the main paper, the measured $\Delta\phi*$ value could have a small correction due to the neutron path integrated magnetic dipole field of the polarized gas, causing an extra precession signal in the neutron phase that is also proportional to $P_3$.  This effect was discussed in Ref.~\cite{Zimmer2002}, however that work did not account for the scalar-contact term that one must include for real particles~\cite{Romalis14,Limes19}, which is different than one would expect from the simple classical result where it is assumed that one can have non-interacting and overlapping point-like particles. For n-$^3$He the contact interaction should be 0 for any reasonable first order approximations, so the neutrons do not experience the classical field of the individual polarized $^3$He nuclei inside the cell, but only the integral of the long range dipole fields caused by a non-spherical geometry inside the polarized cell \cite{Limes19} and, since the neutron precession also integrates the field along their entire flight path, the classical field of a polarized (magnetized) volume outside of the cell.

The scalar contact term is $2\mu_0\delta^3({\bf r}){\bf m}/3$ for non-interacting point dipoles \cite{Vlassenbroek96} leading to a field of $2/3 \mu_oM$ where we replace $M_3$ for $^3$He specifc arguments by the magnetization $M$. For this reason a parameter called $\kappa$ is introduced to represent the actual magnitude of the contact interaction between dissimilar particles as a ratio to that of the ``classical field". Thus, $\kappa>0$ is often referred to as an enhancement factor, such as the case for alkali-metal electrons with a noble gas (e.g. during SEOP) that experience electron paramagnetic resonance shifts much larger than ``expected" from the classical field due to the contact interaction \cite{Romalis98}. For nuclei (or nucleons), the direct contact interaction is 0, but indirect $J$ couplings can still cause a (small) finite effect \cite{Limes19}. 

The field external to the polarized gas volume retains the classical result for a magnetized object; however, as described in Ref. \cite{Limes19} the apparent field experienced by spins inside the volume due to the dipole fields and the contact interactions can be parameterized as
\begin{equation}
    B'_{\rm inside}=\mu_0\left[ M-\eta M+{\frac{2}{3}}(\kappa-1)M\right],
\label{field1}
\end{equation}

\noindent where here we need not worry about a field average as the neutron will sample a line integral through the cell. Here $M$ is the magnetization of the noble gas nuclei, $\kappa$ discussed above describes the spin-spin interaction, and $\eta$ is referred to as the geometric demagnetization factor, i.e. this term parametrizes the modification of the field inside a magnetized volume due to its shape. The first term in Eq.~\ref{field1} is the magnetization of the sample, the second term is the $H$ from classical arguments, and the last term is the quantum mechanics-required correction of the scalar contact interaction.  One can see here that for $\kappa=1$, i.e. the case with perfectly non-interacting particles, one obtains the classical result. For our work, given that any possible $^3$He-neutron coupling will lead to $\kappa=0$ in the first order (which will be sufficient for this correction factor), assuming the classical limit would result in an over-correction.  

The values of $\eta$ have been calculated for normal geometries used in polarization studies such as cylinders \cite{Chen2006}; these values can be solved for analytically in many cases or numerically.  The most recognizable ``demagnetization" factor from classical electrodynamics  is $\eta=1/3$ for a sphere, which allows one to immediately obtain the ``classical field" inside a sphere, $B=2/3\mu_0M$ for example. Here as an additional note, for the case of like particles where $\kappa=0$, by definition we see that $B'_{\rm inside}=0$; this is the reason why a magnetized sphere of hyperpolarized noble gas nuclei do not experience NMR frequency shifts due to their own magnetization, whereas the alkali-metal electrons notably not only experience EPR frequency shifts due to the polarized noble gas, but a much larger shift increased by $\kappa$ due to the enhanced probability of the alkali metal electron being located at the noble gas nucleus in these systems~\cite{Romalis98, Babcock2005}.

If the neutrons only experienced the field inside the cell, we would need to calculate the ``demagnetization" factor ({\it i.e.} edge effects) or $H$ for the particular cell geometry in question; however, the spin precession of the neutrons samples the cell's field over the entire neutron flight path, including outside the cell. It can be shown that for a (classical spherical) dipole and thus an assembly of arbitrary dipoles that: 
\begin{equation}
    \int_{\infty}^\infty H_{dipole}(z)dz=\left({\frac{1}{3}} -{\frac{1}{3}}\right)M=0,
\end{equation}
where $z$ is the axis of the dipole, i.e the direction of magnetization, and also the direction of the neutron flight path in works such as Ref.~\cite{Zimmer2002}. Here $1/3\mu_0M$ is the integral outside the sphere and $-1/3\mu_0M$ is the well-known value of $H$ inside a spherical dipole. Thus for work such as Zimmer et al. \cite{Zimmer2002}, where the ``self" field of the cell's dipole is aligned with the direction of neutron propagation, for a spherical reference volume one quickly realizes that since outside the volume $B=\mu_0H$ and inside the volume the effective field is described by Eq.~\ref{field1} with $\kappa=0$, the additional precession experienced by the neutrons for the case of will be 
\begin{equation}
    \phi^{M_\parallel}={\frac{1}{3}}\mu_0Ml\frac{\gamma_n}{v},
\end{equation} 
where $\gamma_n=2.98\times 10^7$ Hz/T is the gyromagnetic ratio of the neutron, $v$ is the neutron velocity and $l$ the neutron path length through the cell. Since we integrate over the whole neutron flight path, this result will also hold for an assembly of dipoles in the shape of the $^3$He cell. Therefore the data in that work was over-corrected by $2/3\mu_0Ml\gamma_n$, the amount of the reduction in the apparent field, $B'$, from the scalar contact term. This would raise their result by $\approx 0.15\%$. It is interesting to note that this result is analogous to the result for the NMR frequency of an infinitely long cylinder parallel to the $B_0$ field given in eq. 11 in the main paper where $\eta=0$ \cite{Vlassenbroek96}. Here we recover the same result as that for an infinitely long sample, {\it i.e.} the solenoid-like field of a uniformly magnetized rod minus the scalar contact term, because the full integrated field now also includes the field outside of the cell as sampled by the neutrons.

For our geometry the case is different: we have a magnetized cylinder where the cylinder axis, parallel to the neutron flight path, is perpendicular to the holding field $B_0$ and the gas magnetization. For consistency we will leave ``z" as the direction of the neutron flight path and use ``y" as the direction of the noble gas polarization. Here the integral of $H$ from $-\infty \rightarrow +\infty$ becomes:
\begin{equation}
    \int_{\infty}^\infty H_{dipole}(y)dz=-\left({\frac{1}{6}}+{\frac{1}{3}}\right)M=-\frac{1}{2}M.
\end{equation}
From the usual expression of a dipole field one sees that on the equatorial plane, the field direction external to a dipole is reversed and half the magnitude for any distance. Thus the integral along the orthogonal path to the dipole is  $-1/6\mu_0M$ and the additional precession experienced by the neutrons is 
\begin{equation}
    \phi^{M_\perp}=-{\frac{1}{6}}\mu_0Ml{\frac{\gamma_n}{v}},  
\end{equation}

again assuming that $\kappa_0\simeq0$.
Like the prior result, for the case of cell magnetization perpendicular to the $B_0$ field we recover the relation for an infinitely long sample as in eq. 12 of the main paper for which $\eta=1/2$ \cite{Vlassenbroek96}, again because the neutrons also sample the field outside the cell (the - is defined by convention in eq. \ref{field1} for the demagnetization). Substituting in values for the $P_3$ and $N_3$ of the work here this correction would be +0.068\% and is over an order of magnitude below our statistical errors and would only be relevant for 0.1\% or better measurements of $b_i$. For the work of Ref.~\cite{Zimmer2002} this interpretation would raise their cited value by $0.15\%$ as the reported value was over-corrected by $2/3 \mu_0 M_3$, the amount of the scalar contact term.

For the work in the main paper this correction would be $<0.1\%$. However the effect is clearly of importance to precision measurements with nonzero spin particle beams through polarized targets and should not be {\it de-facto} neglected.

\end{document}